\documentclass[copyright,creativecommons]{eptcs}
\providecommand{\event}{FVAV 2017} 
\usepackage{underscore}           

\usepackage{t1enc}
\usepackage[utf8]{inputenc}

\usepackage{amsmath}
\usepackage{amsfonts}
\usepackage{amssymb}
\usepackage{amsthm}
\usepackage{hyperref}
\usepackage{mathtools}

\usepackage{ae,aecompl}
\usepackage{paralist}

\usepackage{calc}
\usepackage{epsfig}
\usepackage{graphicx}
\usepackage{framed}
\usepackage{stmaryrd}
\usepackage{xcolor}
\usepackage{bussproofs}
\usepackage{wrapfig}
\usepackage{yhmath}
\usepackage{enumitem}
\usepackage{tikz}
\usetikzlibrary{arrows,shapes,automata,petri,backgrounds,calc,intersections,decorations.pathreplacing}

\usepackage{upgreek}
\definecolor{myblue}{rgb}{0.2, 0.2, 0.6}
\newcommand{\Blue}{\color{myblue!100}}
\definecolor{myorange}{rgb}{0.91, 0.41, 0.17}
\newcommand{\Orange}{\color{myorange!100}}

\usepackage{macros/roadlogic}
\usepackage{macros/proofs}
\usepackage{macros/autos}
\usepackage{macros/shortcuts}

\title{Imperfect Knowledge in \\ Autonomous Urban Traffic Manoeuvres\thanks{This research was partially supported by the German Research Foundation (DFG) in the Research Training Group GRK 1765 SCARE.}}

\author{Maike Schwammberger
\institute{Department of Computing Science, University of Oldenburg\\ Oldenburg, Germany\vspace{-0.2cm}}\\
\email{schwammberger@informatik.uni-oldenburg.de}
}

\def\titlerunning{Imperfect Knowledge in Urban Traffic Manoeuvres}
\def\authorrunning{M. Schwammberger}

\hypersetup{
	pdftitle={\@title},
	pdfauthor={\@author}
}

\begin{document}
\maketitle
\begin{abstract}
Urban Multi-lane Spatial Logic (UMLSL) was introduced in \cite{HS16} for proving safety (collision freedom) in autonomous urban traffic manoeuvres with perfect knowledge. We now consider a concept of imperfect knowledge, where cars have less information about other cars. To this end, we introduce the concept of a multi-view and propose crossing controllers using broadcast communication with data constraints for turning manoeuvres at intersections.
\smallskip

\end{abstract}
\section{Introduction}\label{sec:introduction}
In urban traffic, turning at an intersection is a challenge for autonomous cars, since other cars approach the crossing from various directions. To pass the intersection, the cars use possibly overlapping parts of the critical resource: the intersection. In a previous paper \cite{HS16}, we proposed \emph{crossing controllers}, that can safely conduct turn manoeuvres with \emph{perfect knowledge}. Here, perfect knowledge means that every car knows the physical size and braking distance of all other cars on the road. In our meaning, \emph{safety} means collision freedom and thus reasoning about car dynamics and spatial properties.

An approach to separate the car dynamics from the spatial considerations and thereby simplify reasoning, was introduced in \cite{HLOR11} with the \emph{\MLSLL{}} (\MLSL) for expressing spatial properties on multi-lane motorways. This logic and its dedicated abstract model was extended with length measurement in \cite{HLO13} for country roads with oncoming traffic. We again extended this approach in \cite{HS16} by introducing a generic topology of \emph{urban traffic networks} and \emph{\UMLSLL{}} (\UMLSL) for reasoning about traffic situations at intersections.

The key contribution of our paper is the adaption of the existing controller for perfect knowledge from \cite{HS16} to a communicating crossing controller with \emph{imperfect knowledge}, meaning that, besides its own braking distance, a car only perceives the physical size of other cars. To cope with this penalty, we extend the crossing controller by a concept of \emph{broadcast communication with data constraints} to communicate with helper controllers which are located in other cars. We also define a \emph{multi-view} covering all roads that meet at an intersection.

Our approach differs from the work of Ody \cite{O17} on monitoring of traffic situations, where the author also uses the abstract model and MLSL from \cite{HLOR11}. There for single sequences of traffic snapshots it is automatically checked if a \MLSL{} formula holds globally throughout the sequence.

We construct our crossing controller based on the design of the controllers in \cite{HLOR11, HLO13, HS16} and specifically for our urban traffic use case. Another approach is to synthesize controllers from given properties, which was already investigated for basic \MLSL{} for highway traffic in \cite{BHLO17}. However, their synthesized controllers abstract from a continuous time dimension.

We consider fully autonomous cars and thus do not model human drivers. Additionally, we do not consider cases where people invade the safety envelope of a car, but refer the approach of Althoff and Magdici~\cite{AM16} for this. There the authors compute an over-approximation of possible occupancies of traffic participants over time to ensure safety of autonomous cars. A different attempt to broaden the approach with \MLSL{} for highway traffic and country roads to intersection scenarios was introduced by Xu and Li in \cite{XL16}. Instead of our directed graph topology, the authors introduce a \emph{space grid model}, where single grids may belong to horizontal lanes or vertical lanes or to no lane at all, e.g. because they are blocked by a building. The authors only apply their results to T-junctions and construct a controller for this special case. Moreover, for transitions between different \traffics s only a discrete time dimension is applied. Loos and Platzer investigate intersections of single lanes with one car on each lane in \cite{LP11}. They use traffic lights as a control mechanism, where a car is not permitted to enter an intersection when the light is red. They verify safety of their hybrid systems with the tool KeYmaera.

This paper is structured as follows. In Sect.~\ref{sec:abstractmodel}, we adapt the abstract model from \cite{HS16} and focus on our extension to the concept of a \emph{multi-view}. We introduce the broadcast communication with data constraints in Sect.~\ref{sec:communication}. We introduce syntax and semantics of the crossing controllers with communication in Sect.~\ref{sec:controllers}, where we also introduce the concept of the crossing and helper controller for imperfect knowledge. We then construct the new crossing controller and its helper controllers. A conclusion, some further related work and ideas for proving safety of our controllers in future work are given in Sect.~\ref{sec:conclusion}.

\section{Abstract Model}\label{sec:abstractmodel}
We start with an informal introduction of the considered abstract model for urban traffic scenarios and give more formal details for central concepts in the respective subsections. Topics marked with \mystar{} are from our previous paper \cite{HS16} and only introduced briefly, while their formal definition from \cite{HS16} can also be found in the appendix.

The abstract model contains a set $\bigcrossingsegment$ of \emph{crossing segments} $c_0 , c_1 , \ldots$ and a set $\Lanes$ of \emph{lanes (lane segments)} $0, 1, \ldots$ connecting different crossings. Each crossing segment and each lane (segment) has a finite length. Adjacent lanes are bundled to \emph{road segments} $\{0,1\}, \{2,3\}, \ldots \in\roadsegment$ such that $\roadsegment$ is a subset of $\mathcal{P}(\Lanes)$. Typical representatives of $\roadsegment$ are $r_0, r_1, r_2, \ldots$. Adjacent crossing segments form an intersection, e.g. named by $cr$. The connections of lane and crossing segments are defined by an underlying graph topology called \emph{urban road network} \net{} (cf. Sect.~\ref{sec:topology}).

\begin{wrapfigure}[16]{r}{0.49\textwidth}
\centering
	\includegraphics[scale=.63]{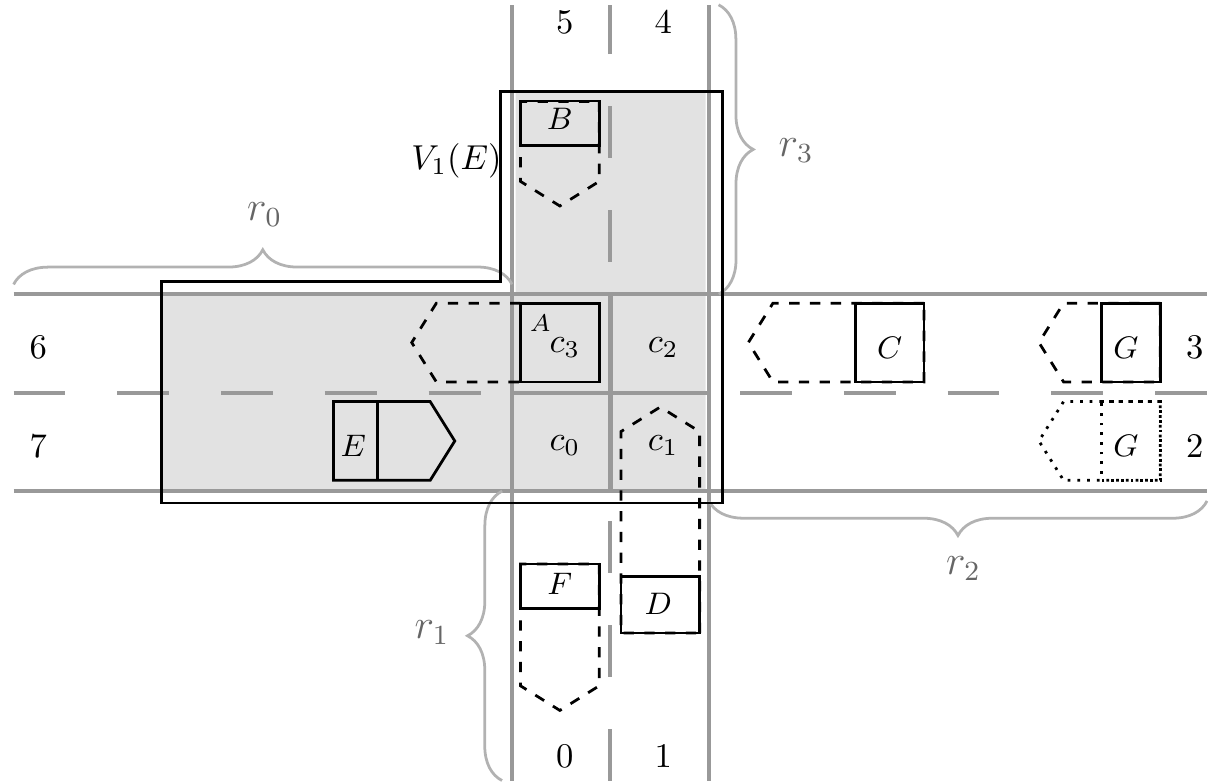}
	\caption[]{Car $E$ perceives the physical size of other cars in its view $V_1 (E)$. The dashed braking distances of other cars are invisible for $E$.}
	\label{fig:trafficsituation2}
\end{wrapfigure}

Every car has a unique \emph{car identifier} $A, B, \ldots$ from the set $\mathbb{I}$ of all car identifiers and a real value for its position $pos$ on a lane. We use car $E$ as the \emph{car under consideration} (short \emph{ego car} or \emph{actor}) and introduce the special constant $ego$ with valuation $\nu (ego) = E$ to refer to this car. While a \emph{reservation} $re(\ego)$ is the space car $E$ is actually occupying, a \emph{claim} $cl(\ego)$ is akin to setting the direction indicator representing the space a car plans to drive on in the future (cf. dotted part of car $G$ in Fig.~\ref{fig:trafficsituation2}, where $G$ plans to change its lane). This static information about cars like position, reservation and claim is captured in a \traffics{} \Road{} (cf. Sect.~\ref{sec:trafficsnapshot}). To simplify reasoning, only local parts of the \traffics{} are considered as every car has its own local \emph{view}, cf. view $V_1 (E)$ of car $E$ in the example (cf. Sect.~\ref{sec:view}). For logical reasoning, we then evaluate formulae of the \emph{\UMLSLL{}} (\UMLSL) over a view (cf. Sect.~\ref{sec:mlsl}).

With \emph{imperfect knowledge}, we assume, that the actor $E$ only perceives those parts of other cars it can perceive with its sensors: The physical position and size of the car (cf. solid parts of cars in Fig. \ref{fig:trafficsituation2}), but not the braking distance (cf. dashed parts). Only ego car $E$ itself knows its own braking distance and thus its whole \emph{safety envelope}, while the braking distances of the other cars are invisible to $E$. In our approach, the safety property is already violated, if a car invades the braking distance of another car and not only if a physical collision occurs. The idea is, that in case of an emergency braking manoeuvre our safety property is still valid.

We distinguish between the movement of cars on lanes and on crossings. We allow for two-way traffic on lanes of continuous space and finite length, where every lane has one direction and cars normally drive on a lane in the direction of increasing real values, but may temporarily drive in the opposite direction for overtaking. As a car's direction changes while turning at an intersection, we can not assign one specific direction to a crossing segment and consider them as discrete and either fully occupied by a car or empty. An example for this is car $A$ in Fig.~\ref{fig:trafficsituation2}, where $A$ occupies the whole discrete crossing segment $c_3$. When a car is about to drive onto a discrete crossing segment and time elapses, the car's safety envelope stretches to the whole crossing segment, while disappearing continuously on the lane it drove on.
\subsection{Topology}\label{sec:topology}
We restrict the abstract model to road segments with two lanes, one in each direction. Intersections are four connected crossing segments with four road segments meeting at a crossing (cf. example in Fig.~\ref{fig:trafficsituation2}). We describe connections between lanes and crossing segments by an \emph{Urban Road Network}\mystar{} $\net$, whose nodes are from the set $\mathrm{V} = \mathbb{L} \cup \bigcrossingsegment$ of lanes and crossing segments. As we are dealing with traffic that is evolving over time, we capture the (finite and real valued) length of lanes and crossing segments in our graph by assigning a weight $\omega(v)$ to each node $v \in\mathrm{V}$. Adjacent crossing segments form strongly connected components $\mathit{I}_{cs}$ (intersections, abbreviated with $cr$). Neighbouring lanes, connected with an undirected edge for bidirectional lane change manoeuvres, are components $\mathit{I}_{l}$ (road segments, abbreviated with $r$). Edges from the sets $\mathbb{L} \times \bigcrossingsegment$, $\bigcrossingsegment \times \mathbb{L}$ and $\bigcrossingsegment \times \bigcrossingsegment$ are directed, whereby entry and exit points to the intersection are defined unambiguously. With these connected components, we can construct a coarser version ${\net}_{\mathit{I}}$ of $\net$, where a road segment $r$ is connected with a crossing $cr$ with a directed edge $(r,cr)$ resp. $(cr,r)$, iff there exists a matching directed edge in the underlying graph $\net$.

\begin{figure}[htbp]
    \centering
    \includegraphics[scale=.8]{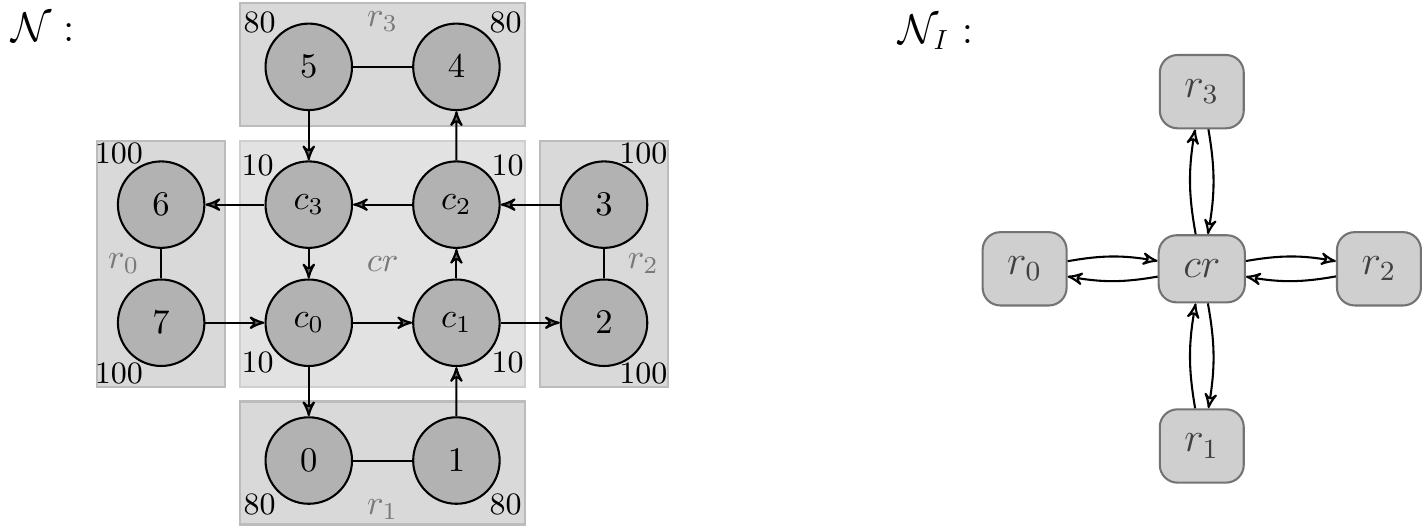}
    \caption[]{Urban road network \net{} corresponding to Fig. \ref{fig:trafficsituation2} left and coarser version ${\net}_{\mathit{I}}$, only depicting the strongly connected components and their relations with each other, at the right.}
    \label{fig:roadnetwork1}
\end{figure}

The corresponding road network $\net$ to Fig.~\ref{fig:trafficsituation2} is depicted on the left side of Fig.~\ref{fig:roadnetwork1} and the coarser version ${\net}_{\mathit{I}}$ on the right. A suitable \emph{path} for car $E$ in \net{} is $\pth(E) = \langle \ldots, 7 , c_0 , c_1 , c_2 , 4, \ldots \rangle$, where it plans on turning left. This \emph{fine-grained path} is used later to determine the parts of lanes and crossing segments an arbitrary car occupies in a view.

The coarser version of this path is ${\pth(E)}_{\mathit{I}} = \langle \ldots, {r}_0 , cr , {r}_3 , \ldots \rangle$, where $cr$ is the name of the whole intersection. Such \emph{coarse-grained paths} are used later to build the virtual lanes for our multi-view in Sect.~\ref{sec:view}.

\subsection{Traffic Snapshot}\label{sec:trafficsnapshot}
A \emph{traffic snapshot}\mystar{} $\Road$ captures the traffic on an \roadnetwork{} \net{} at a given point in time and is defined by the structure
\begin{align*}
 \RoadDef\text{,}
\end{align*}
where $\pth(C)$ is the path an arbitrary car $C$ traverses in the urban road network \net. The index $\curr(C)$ relates to the node $\pth(C)_{\curr(C)}$ the car $C$ is currently driving on. With $pos(C)$ the real-valued position of the rear of car $C$ on $\pth(C)_{\curr(C)}$ is defined. The set $clm(C)$ (resp. $res(C)$) is the set of all lanes $C$ claims (resp. reserves) and $cclm(C)$ (resp. $cres(C)$) is the set of crossing segments $C$ claims (resp. reserves). Amongst others, we demand the following \emph{sanity conditions} to hold for an arbitrary traffic snapshot \Road:
\begin{align}
	&0 \leq |res(C)| \leq 2, &\hspace{0.0cm}& 0 \leq |clm(C)| \leq 1, \hspace{1.5cm} |res(C)| + |clm(C)| \leq 2, \label{sanity1}\\
	&1 \leq |res(C)| + |cres(C)|, &\hspace{0.0cm}& |cclm(C)| \geq 1 \rightarrow |clm(C)| = 0 \wedge |res(C)| = 1\text{.}\label{sanity2}
\end{align}
With conditions \ref{sanity1}, we only allow for one lane change manoeuvre at once, where either with $|clm(C)|=1$ a lane is claimed or with $|res(C)|=2$ the car is already changing lanes. With conditions \ref{sanity2}, we state that at any point in time, a car reserves at least one lane or crossing segment and we only allow for a crossing claim, if car $C$ is not involved in a lane change manoeuvre. Thus, a car may not enter an intersection with an active lane change manoeuvre.

To model the behaviour of cars, we allow \emph{evolution transitions}\mystar{} between traffic snapshots which respect the sanity conditions. The node in $\pth(C)$ that is reached after some time $t$ elapses, we call $\pth(C)_{\nexxt(C)}$. This node can either be a crossing or lane segment. Note that $\pth(C)_{\nexxt(C)}$ is the node, where after $t$ time units the position of the rear of car $C$ is located, while it is possible, that the safety envelope of $C$ stretches to more nodes. When approaching an intersection $cr$, we claim all needed crossing segments from $cr$ that car $C$ traverses in its path $pth(C)$ at once. Note that $\curr(C) = next(C)$, iff $C$ did not move far enough to leave its current node.
\subsection{Imperfect Knowledge and Multi-View}\label{sec:view}
For logical reasoning we consider only finite parts of the traffic snapshot \Road. The idea is that the safety of a car depends only on its immediate surroundings. We therefore use the concept of a local \emph{view}, which only contains those parts of lanes and crossing segments that are within some \emph{horizon} $h$ around the actor $E$. In previous work \cite{HLOR11,HLO13} covering highway and country road traffic, the set of lanes $L$ in a view was obtained by taking a subinterval of the global set of parallel lanes $\mathbb{L}$. This is no longer possible for the urban traffic scenario, since taking an arbitrary subinterval of lanes can yield a set of lanes which are not connected. We therefore construct a view from the urban road network, the current traffic snapshot, a given real-valued interval $X=[a, b]$ and the owner of the view $E$.

\begin{definition}[View]
For a road network \net{} and its nodes $\mathrm{V}$ the \emph{view} $V(E)=(L,X,E)$ of car $E$ contains a set of \emph{virtual} lanes $L\subseteq {\cal P}({\mathrm{V}}^{\Z})$, an interval of space along the lanes $X=[a,b]\subseteq\R$ visible in $V(E)$ and $E\in\mathbb{I}$ as the car identifier of car $E$ under consideration.
\end{definition}

If an intersection is within the horizon $h$, we deal with a bended view as cars are allowed to turn in any possible direction at the crossing (cf. view $V_1(E)$ in Fig.~\ref{fig:trafficsituation2}). To allow for spatial reasoning with our logic \UMLSL, we flatten the view by constructing a straight \emph{virtual view} from the urban road network $\mathcal{N}$ and the path $\pth(E)$ of car $E$. As we currently only consider intersections of two by two lanes, one virtual view is also composed of two \emph{virtual lanes}.

For perfect knowledge, it was sufficient to consider only that virtual view for the actor $E$ which corresponds to its path $\pth(E)$ (cf. view $V_1(E)$ in Fig.~\ref{fig:trafficsituation2}, where $E$ plans on turning left). With imperfect knowledge, $E$ can not perceive whether the safety envelope of a car that is not (yet) physically driving on the crossing already stretches to some crossing segments.

\begin{figure}[htbp]
\centering
	\includegraphics[scale=.74]{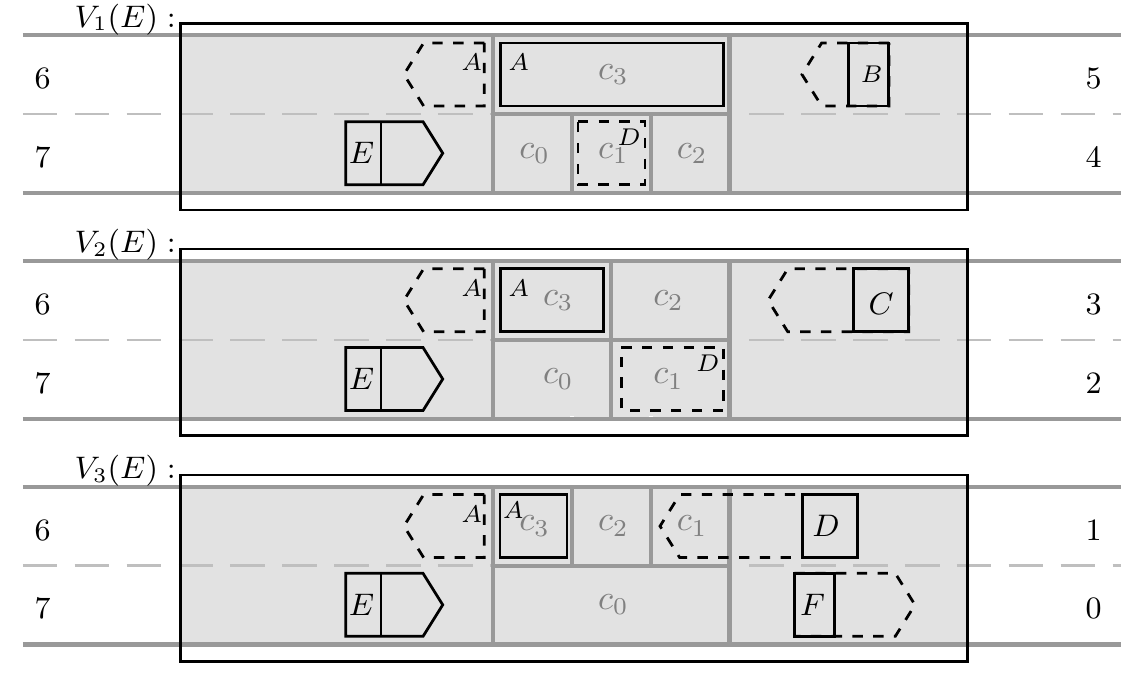}
	\caption[]{The virtual multi-view $V_m (E) = ( V_1 (E) , V_2 (E) , V_3 (E) )$ of car $E$ covers the road segment $E$ is driving on, the intersection and all other road segments linked to the intersection.}
	\label{fig:virtualview}
\end{figure}

Consider again the example from Fig.~\ref{fig:trafficsituation2}, where car $E$ does not perceive the braking distance of car $D$ which already stretches to the intersection. To cope with the imperfect knowledge, we propose, that car $E$ communicates with all cars on the intersection and with all cars that are approaching the intersection from any direction. Therefore, we need to consider more than the previously introduced one bended view $V_1 (E)$ and introduce the concept of a virtual \emph{multi-view} $V_m (E)$. This view covers the already introduced view $V_1 (E)$ as well as view $V_2 (E)$ covering road segment $r_0$, the intersection and segment $r_2$ and view $V_3 (E)$, covering $r_0$, the intersection and $r_1$. Note that we do not consider the u-turn direction of the intersection, because $r_0$ is already covered in all other virtual views. The constructed multi-view is depicted in Fig.~\ref{fig:virtualview}.

To formally build the multi-view, we first identify the road segment $E$ is currently driving on with the underlying graph topology (cf. Sect.~\ref{sec:topology}). As in Sect.~\ref{sec:trafficsnapshot}, the current path segment $E$ is driving on is defined by $\pth(E)_{\curr(E)}$ (in the example: $7$) and the related road segment $r_{curr(E)}$ is given through the strongly connected component $\mathit{I}_{l} (\pth(E)_{\curr(E)})$ (in the example: $\mathit{I}_{l} (7) = r_0$). When a crossing is ahead, the first crossing segment $E$ will drive on when entering the intersection is given by $\pth(E)_{next(E)}$ (in the example: $c_0$) and therefore the whole intersection $cr$ is obtained through the connected component $\mathit{I}_{cs} (\pth(E)_{next(E)})$ (in the example: $\mathit{I}_{cs} (c_0) = cr$).

Next we identify all road segments ${r}_i$ apart from $r_{curr(E)}$ which are connected to the intersection $cr$ with a directed edge $(r_i ,cr)$ in the coarser graph ${\net}_{\mathit{I}}$. We can simply do this by considering all crossing segments $c_i \in cr$ and identifying the lanes $l_i \in {r}_i$ which have a directed edge to $c_i$ in \net. This way, we detect all road segments from which cars can enter the junction and derive pairs of virtual lanes, later needed for the construction of the respective virtual views.

\begin{definition}[Virtual Lanes]\label{def:vlanes}
Consider a car $E$, its current path element ${\pi}_i = \pth(E)_{\curr(E)}$ and its next path element $\pi_{i+1} = \pth(E)_{next(E)}$. We derive the neighbouring lane ${\pi}_{i,n}$ to ${\pi}_i$ from the urban network $\mathcal{N}$, where it is the only node connected to ${\pi}_i$ with an undirected edge. The current road segment is defined by $r_{curr(E)} := \mathit{I}_{l} (\pi_i)$ and the next intersection by $cr := \mathit{I}_{cs} (\pi_{i+1})$.

We use the function $pre(cr)$ to identify the set of all predecessor nodes $r_j$ with an edge $(r_j , cr)$ in the coarser graph ${\net}_{\mathit{I}}$. The \emph{coarser virtual lanes} $\mathbf{L}_j$ are given through
\begin{align*}
    \forall r_j \in pre(cr) \wedge r_j \neq r_{curr(E)} : \mathbf{L}_j = (r_{curr(E)} , cr , r_j) \text{.}
\end{align*}

To identify the corresponding \emph{finer virtual lanes} $\overrightarrow{{\pi}_j}$ (driving direction according to $E$'s driving direction) and $\overleftarrow{{\pi}_j}$ (driving direction opposite to $E$'s driving direction) contained in each $\mathbf{L}_j$, we identify the shortest directed path forwards from ${\pi}_i$ to an element ${{\pi}_{j_1}} \in r_j$ to build $\overrightarrow{{\pi}_j}$. For $\overleftarrow{{\pi}_j}$, we search the shortest directed path backwards from an element ${\pi}_{j_2} \in r_j$ to the neighbouring node ${\pi}_{i,n}$. We then derive for all coarser virtual lanes $\mathbf{L}_j$ the finer virtual lanes
\begin{align*}
    \overrightarrow{{\pi}_j} = [{\pi}_i , \overrightarrow{cs} , {\pi}_{j_1} ] \text{ and } \overleftarrow{{\pi}_j} = [{\pi}_{i,n} , \overleftarrow{cs} , {\pi}_{j_2} ]\text{,}
\end{align*}
where $\overrightarrow{cs}$ and $\overleftarrow{cs}$ are the respective shortest directed subpaths through the intersection $cr$. The set of all \emph{virtual lanes} is given by $L_m = \{ (\overrightarrow{{\pi}_1}, \overleftarrow{{\pi}_1}) , \ldots , (\overrightarrow{{\pi}_n}, \overleftarrow{{\pi}_n}) \}$, where $n:= |pre(cr)| - 1$.
\end{definition}

From definition \ref{def:vlanes} we obtained pairs $L_j$ of virtual lanes $\overrightarrow{\pi_i}$ and $\overleftarrow{\pi_i}$, which each are used to build one virtual view $V_j (E, \Road) = (L_j , X , E)$ for a \traffics{} \Road. All virtual views together lead to multi-view $V_m (E, \Road) = ( V_1 (E, \Road) , \ldots , \V_n (E, \Road) )$. To build these virtual views $V_j (E, \Road)$ from the virtual lanes $L_j$, we need to define the size of the extension $X=[a,b]$ along the lanes. From the position $pos(E)$ of the car under consideration $E$, we look forwards and backwards up to a sufficient constant horizon $h_f$ resp. $h_b$. We make sure that $h_f$ is big enough, that a fast car approaching the intersection, that can already have a crossing claim or reservation on the intersection, is included in ${h_f}$\mystar. We consider the same extension $X=[pos(E)- h_b , pos(E) + h_f]$ for each pair of virtual lanes. A virtual view is then defined from the pairs of virtual lanes with the described extension as follows.

\begin{definition}[Virtual view and multi-view]\label{def:multiview}
For a car $E$, a \traffics{} \Road, a pair of virtual lanes $L_i = (\overrightarrow{{\pi}_i}, \overleftarrow{{\pi}_i})$ and the extension $X=[pos(E) - h_b , pos(E) + h_f]$ the \emph{virtual view} $V_i$ of $E$ is defined by
$V_i (E, \Road) =(L_i , X, E)$.

The set of all virtual views for car $E$, built for one intersection $cr$ is named the \emph{multi-view} $V_m (E,\Road) = (V_1 (E, \Road), \ldots , V_n (E,\Road))$, where $n$ is the amount of pairs of virtual lanes $L_i$ constructed through Def.~\ref{def:vlanes}. We abbreviate $V (E, \Road) = V(E)$ if \Road{} is clear from context.
\end{definition}

\noindent
\textbf{Sensor Function.} The car dependent sensor function $\Omega_E:\ID\times\Roads\rightarrow \R$ yields, given an arbitrary car $C$ and a traffic snapshot $\Road$ the physical size of a car $C$ as perceived by $E$'s sensors.

\noindent
\textbf{Visible Segments of Cars in a View\mystar.} For both virtual lanes, we need to find all segments $seg_V(C)$ which are (partially) occupied by a car $C$ and visible in the view of $E$. Considering highway traffic on continuous lanes, it is easy for a car $E$ to obtain the interval of space $[a_c , b_c ]$ another car $C$ occupies in its view $V(E)$ through $a_c := pos(C)$ and $b_c := pos(C) + \Omega_E (C)$.

For urban traffic with intersections, it is a lot more complicated to address this task, because lanes as well as crossing segments are of finite length. Thus, the perceived size $\Omega_E (C)$ of a car $C$ may stretch over several (connected) lane and crossing segments in the road network \net{} (cf. car $A$ in Fig.~\ref{fig:trafficsituation2}, whose physical part will occupy crossing segment $c_3$ and a part of lane segment $6$ when it leaves the intersection in the near future). We therefore construct the set of segments $seg_V(C)$ another car $C$ occupies in the virtual view of car $E$ by taking the position of $C$, its size $\Omega_E (C)$ and the weight of nodes as defined in the road network $\net$ into account. For details for the construction of $seg_V(C)$, we refer to \cite{HS16}.
\subsection{\UMLSLL}\label{sec:mlsl}
Using car variables $c \in \carvariables{} \cup \{\ego\}$ ranging over car identifiers and variables $u,v \in \carvariables{}\cup\realvariables{}$ with \realvariables{} ranging over the real numbers the syntax of \UMLSL{} formulae is defined by
  \begin{align*}
    \phi &::= \true\mid u=v\mid \free\mid cs\mid \reserved{c} \mid
    \claimed{c} \mid \lnot \phi \mid \phi_1 \land \phi_2 \mid \exists c \qsep \phi_1
\mid \phi_1 \chop \phi_2 \mid {}_{\phi_1}^{\phi_2}\text{.}
  \end{align*}
We use the atom $\free$ to represent free space and $cs$ for crossing segments. Hereby, we can e.g. state that car $E$ claims ($cl(\ego)$) or reserves ($re(\ego)$) a crossing segment ($cs \wedge (cl(ego) \vee re(\ego))$) or that a crossing segment is free ($cs \wedge \free$). We can formalise the size of a horizontal interval in \UMLSL, where e.g. $\free \wedge \ell > d$ holds, if there is an interval of free space on a lane exceeding the size $d \in \mathbb{R}^+$. Besides these atoms, Boolean connectors and first-order quantifiers, formulae of \UMLSL{} use two \emph{chop operators}. One for a horizontal chop, denoted by \(\phi_1\chop\phi_2\) like for interval temporal logic \cite{Mos85} and one for a vertical chop given by the vertical arrangement of formulae ${}_{\phi_1}^{\phi_2}$. Intuitively, a formula \(\phi_1\chop\phi_2\) holds if we can split the view \(V\) horizontally into two views \(V_1\) and \(V_2\) such that on \(V_1\) \(\phi_1\) holds and \(V_2\) satisfies \(\phi_2\). Similarly a formula ${}_{\phi_1}^{\phi_2}$ is satisfied by \(V\), if \(V\) can be chopped at a lane into two subviews, \(V_1\) and \(V_2\), where \(V_i\) satisfies \(\phi_i\) for \(i=1,2\). 

In a part of view $V_1 (E)$ (cf. Fig.~\ref{fig:trafficsituation2}) the formula $\phi \equiv re(\ego) \chop \free \chop cs \wedge \free$ holds. Here, $re(\ego)$ is the space car $E$ reserves on lane $7$, the atom $\free$ represents the free space in front of car $E$, and $cs \wedge \free$ stands for the unoccupied space on crossing segment $c_0$.

In case of a single (possibly virtual) view $V(E)$ of car $E$, the semantics\mystar{} of \UMLSL{} formulae is evaluated over a \traffics{} \Road, the view $V(E)$ and a valuation $\nu$, which defines the current valuation $\nu(u)$ of variables $u$ with elements from $Var = \mathbb{I} \cup \mathbb{R} \cup \mathbb{CS}$. In case of a multi-view $V_m$, we define the following satisfaction of a formula $\phi$ over $V_m$.

\begin{definition}[Multi-view semantics of \UMLSL{} formulae]\label{def:semantics-multiview}
 For a multi-view $V_m = \{ V_0 ,\ldots, V_n \}$, a \traffics{} \Road{} and a valuation $\nu$ the \emph{satisfaction} of a formula $\phi$ is defined by
\begin{align*}
  \Road, V_m , \nu\models\phi \;\Leftrightarrow\; \forall V_i \in V_m : \Road, V_i , \nu\models\phi \text{.}
\end{align*}
Existential satisfaction over a multi-view is possible with $\exists V_i \in V_m : \Road, V_i , \nu\models\phi$.
\end{definition}
\noindent
\textit{Abbreviations.} We use the abbreviation $\langle \phi \rangle$ to state that a formula $\phi$ holds \emph{somewhere} in the considered view. We use abbreviations like ${\phi}^{< d}$ or ${\phi}^{> d}$ for $\phi \wedge \ell < d$ resp. $\phi \wedge \ell > d$.

\noindent
\textbf{Twisted views and the evaluation of \UMLSL{} formulae.}
For highway traffic and country roads \cite{HLOR11,HLO13}, spatial formulae of MLSL are evaluated from ``left to right''. In urban traffic, a car $C$ builds up the virtual multi-view from its own perspective, to evaluate formulae of the \UMLSL. Consider again Fig.~\ref{fig:virtualview}. In view $V_1 (E)$, the formula $\phi \equiv \langle re(E) \chop \free \chop cs \rangle$ holds. Now consider the respective view $V_1 (B)$, comprising the same lane and crossing segments as $V_1 (E)$, but build up from the sight of car $B$. This view is comparable with $V_1 (E)$, twisted around by $180$ degrees. In view $V_1 (B)$, the formula $\phi \equiv \langle re(E) \chop \free \chop cs \rangle$ does not hold, whereby its inverse version ${\phi}^{-1} \equiv \langle cs \chop \free \chop re(E) \rangle$ holds.
\section{Broadcast Communication with Data Constraints}\label{sec:communication}
In our abstract model, the autonomous cars can be understood as nodes in a \emph{Vehicular ad-hoc network} (VANET), without a fixed wireless infrastructure and without taking roadside units into account. In \cite{OS17}, we proposed a concept of broadcast communication with data constraints for the there introduced hazard warning controllers. We reuse this communication concept for the controllers we introduce in Sect.~\ref{sec:controllers} and which are modelled as extended timed automata \cite{AD94}. One extension is the use of data variables and data constraints in guards, invariants and variable updates, as described by Behrmann et al. in \cite{BDL04} for UPPAAL. We broaden this use of data constraints in timed automata even more by sending data via broadcast channels.

Alrahman et al. propose a \emph{Calculus for Attribute-based Communication} in~\cite{AN15}. The authors consider systems with a large amount of dynamically adjusting components that interact via broadcast channels. Components broadcast valuations of data variables $u$ via an attribute-based output $(u)@\Uppi$ to all processes whose attributes satisfy the predicate $\Uppi$. By using updates $a:=u$ of local attributes $a$, the received data $u$ can be used locally by these processes. Other components only then synchronise with an output $(u)@\Uppi$ when they have an input $\Uppi(x)$ and their local attributes $a$, together with the received message $x$, satisfy the predicate $\Uppi$. We adapt this concept of synchronisation in the definition of input and output actions for our controllers.

For data types on our channels, we use the Z notation~\cite{WD96} of sequences: $seq\;X$ is the set of all finite sequences of elements from a given set $X$. A sequence $s$ consisting of elements $A, B, C$ is written as $s = \langle A, B, C \rangle$. It stands for a function $s = \{ 1 \mapsto A, 2 \mapsto B, 3 \mapsto C \}$ from indices $1,2,3$ to elements $A,B,C$. Thus the $i$th element of $s$ is denoted by function application $s(i)$, e.g., $s(2)=B$. The \emph{length} of $s$ is derived by $\# s$, here $\#s = 3$. For the empty sequence $\langle\rangle$ the length is $0$.

\begin{definition}[Input and Output actions]\label{def:inputoutput}
    For a finite list of data variables $d=\langle d_1, \ldots, d_n \rangle$ and a \UMLSL\ formula $\varphi$ we define an \emph{output action} $\OUT$ on a broadcast channel $a$ by $\OUT := a!d$ and a related 
    \emph{input action} $\IN$ by $\IN := a?d: \varphi$. 
    The set of data variables $d_i \in\mathbb{D}$ ranges over the set of all car identifiers $\mathbb{I}$, the power set $\mathcal{P}(\mathbb{L})$ (resp. $\mathcal{P}(\mathbb{CS})$) of the set of all lanes $\mathbb{L}$ (resp. all crossing segments $\mathbb{CS}$), and finite sequences \emph{seq} $\mathbb{I}$, \emph{seq} $\mathbb{L}$ and \emph{seq} $\mathbb{CS}$.
\end{definition}

\noindent
\textbf{Abbreviation.} We abbreviate $\langle d_1 \rangle = d_1$ for a single data variable $d_1$.

\noindent
\textbf{Example.} A request of car $E$ for some crossing segments is sent via broadcast channel $cross$ with the output $cross ! \langle ego,cs_{ego} \rangle$. Here, $cs$ is the set of crossing segments car $E$ claims for its turning manoeuvre and $\nu(\ego)$ is the senders car identifier. Consider a corresponding input $cross? \langle c,cs \rangle: a \neq c \;\wedge\; cs \cap cs_{a} = \emptyset$, where $\nu(a)$ is the car identifier of the request receiving controller and $cs_{a}$ is the set of crossing segments this car reserves or claims itself. The received data is stored by the receiver in local variables: $\nu(c) = \nu(ego)$ and $\nu(cs) = \nu(cs_{ego})$. This input synchronises with the output iff the \UMLSL{} formula $a \neq c \wedge cs \cap cs_{a} = \emptyset$ evaluated over valuation $\nu$ holds.

\section{Controllers for Safe Crossing Manoeuvres}\label{sec:controllers}
In \cite{HS16}, we introduced a \emph{\cc} to perform turn manoeuvres at intersections with perfect knowledge. This controller made driving decisions according to the current view and traffic snapshot, where it was able to perceive the whole safety envelope of other cars and thus had information about all reserved or claimed lanes and crossing segments of other cars. With imperfect knowledge, ego car $E$ is not able to perceive if the braking distance of another car stretches up to the crossing segments $E$ plans to reserve for itself. Therefore, ego car $E$ has to actively communicate with those cars to prevent collisions. For this purpose we adapt the crossing controller for perfect knowledge from \cite{HS16} with broadcast communication elements as introduced in Sect.~\ref{sec:communication} and introduce a helper controller. This helper concept roughly follows the helper approach for imperfect knowledge for highway traffic from \cite{HLOR11}.

In previous works \cite{HLOR11,HLO13,HS16} we showed that if every car is equipped with the respective proposed controllers for the different traffic scenarios, safety in the sense of disjointedness of reservations is preserved under all time and action transitions. We check the property
\begin{align}\label{formula:safe}
\text{\emph{Safe}}(ego) \;\equiv\; \neg \exists c \colon c \neq ego \land \somewhere{\reserved{ego} \land \reserved{c}}
\end{align}
from the viewpoint of ego car $E$ and use the somewhere operator $\langle\rangle$. \emph{Safe}$(ego)$ states, that there is never a spatial overlap of the reservation of $E$ with the reservation of another car. Note that by demanding the disjointedness of (the speed-dependent) reserved spaces, the formula indirectly requires that $E$ lowers its speed (to shorten its reserved space) when a car ahead of it starts breaking. To maintain \emph{Safe-re}$(ego)$ under time transitions, each car has a \emph{distance controller} as proposed by Damm et al. in~\cite{DHO06}. Rizaldi et al.~\cite{RIA16} examine safety distances for autonomous vehicles, which is useful for such a distance controller. For urban traffic we additionally demand that the distance controller keeps a positive distance to an intersection, if the car does not get permission to enter the intersection. In worst case the car comes to a standstill in front of the crossing until permission to conduct its planned turn manoeuvre is granted. The described distance controller initiates acceleration and braking manoeuvres for the car, which means setting inputs for the actuators on a lower level of controllers. A good example for such a controller on the dynamics level is given by Damm et al. in \cite{DMR14}, where the authors introduce a velocity controller. In our approach, we explicitly separate our controllers from these car dynamics level and focus on a decision making level. That is, our controllers, e.g., decide how and whether a lane change or a crossing manoeuvre is conducted. This approach allows for a purely spatial reasoning. However, a link between the spatial and dynamic reasoning is formalised in~\cite{ORWprocos17}.

Road segments between intersections are structurally comparable to country roads, wherefore we refer to \cite{HLO13}, where a lane change controller for these roads was presented. We only modify this \emph{\rc} by the requirement, that as soon as a crossing is ahead within some distance $d_c$, any claim must be withdrawn immediately and no new claim or reservation might be created until the crossing is passed. However, the car may finish an already begun overtaking manoeuvre, wherefore we make sure the distance $d_c$ is big enough to do so. We assume crossings to be at least $d_c$ apart from each other to guarantee correct functionality of our controllers.

\subsection{Automotive-controlling Timed Automata}\label{sec:acta}
In \cite{HS16}, we introduced extended time automata, called \emph{automotive-controlling timed automata} (ACTA)\mystar, to formalise the controllers for different traffic scenarios from \cite{HLOR11,HLO13,HS16}. As variables these controllers use both clock and data variables. For clock variables $x,y \in \mathbb{X}$ and clock updates we refer to the definition of timed automata and for data variables $d_i \in\mathbb{D}$ and data updates we refer to the extension of timed automata proposed for UPPAAL. These clock and data updates ${\nu}_{act}$ are allowed on transitions of the automata. Note that we allow for the same set of data variables $\mathbb{D}$ we introduced in Def.~\ref{def:inputoutput} for input and output actions, including sets and lists.

Further on, the controllers use \UMLSL{} formulae ${\varphi}_{U}$ as well as clock and data constraints ${\varphi}_\mathbb{X}$ resp. ${\varphi}_\mathbb{D}$ as guards $\varphi$ on transitions and as invariants $I(q)$ in states $q$. An example for a data constraint for a variable $l \in Var$ is $l > 1$. We extend the data constraints for single variables from $Var$ by set operations, which e.g. allows for $cs \cap cs' = \emptyset$ as a guard or invariant, where $cs,cs' \in \mathcal{P}(\mathbb{CS})$. The set $\Phi$ of all guards and invariants is defined by $\varphi \:\equiv\; {\varphi}_\mathbb{U} \;|\; {\varphi}_\mathbb{X} \;|\; {\varphi}_\mathbb{D} \;|\; {\varphi}_1 \wedge {\varphi}_2 \;|\; true\text{.}$

We use the broadcast communication as defined in Sect.~\ref{sec:communication}. Remember that we consider output actions $\OUT$ which can synchronise with appropriate input actions $\IN$ in another controller. We also use \emph{controller actions} $c_{act}$ to commit lane change manoeuvres on road segments and turning manoeuvres at crossings, where e.g. \texttt{rc(}$\ego$\texttt{)} is a \emph{crossing reservation} action for ego car $E$ and \texttt{wd rc(}$\ego$\texttt{)} is the respective withdrawal action for a crossing reservation.

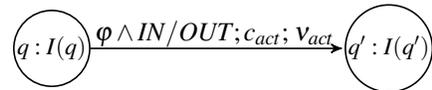
\begin{wrapfigure}[7]{r}{0.35\textwidth}
\centering
	\begin{tikzpicture}[initial text=,->,>=stealth',shorten >=1pt,auto,inner sep=0.5pt,minimum size=0pt,node distance=5.4 cm, semithick, scale = 0.83, transform shape]
	\begin{scope}
		\node [state]	(q0) [xshift=-0.0cm, label=above left:]{$q : I(q)$};
		\node [state]		(q) [xshift=+0.0cm, right of=q0, label=above right:]{$q' : I(q')$};
	    	\path (q0) edge [bend left=0]  node[above] {\large{$\varphi \wedge \IN / \OUT ; c_{act}; \, {\nu}_{act} \;$}} (q);
	\end{scope}
	\end{tikzpicture}
	\caption[]{Syntax elements of an \acta{} with communication}\label{transitionacta}
\end{wrapfigure}
A transition in an ACTA comprises the elements depicted in Fig.~\ref{transitionacta}. The guard $\varphi \land IN$ shown before the separator $/$ has to hold with respect to the current traffic snapshot $\Road$, the view $V (E)$ of ego car $E$ and the valuation $\nu$ in order to execute the output, controller and update actions shown after the separator $/$, yielding a successor state $q'$ and a valuation $\nu'$. The invariant $I(q')$ has to hold in $q'$.

\subsection{Imperfect Knowledge}\label{sec:imperfectknowledge}
In order to enter a crossing, a car first needs to claim a path through the crossing for its turn manoeuvre and check whether there is an overlap of this claim with the claim or reservation of another car, formalised by the \emph{potential collision check}
    \begin{align}\label{formula:pc}
	pc(c) \; \equiv \; c \neq \ego \wedge \langle cl(\ego) \wedge (re(c) \vee cl(c)) \rangle \text{.}
    \end{align}
If a potential collision is detected, the ego car must withdraw its claim. However, with imperfect knowledge the ego car is not able to detect a potential collision with the whole safety envelope of another car, but only with its physical size. Therefore, ego car $E$ has to communicate with cars that might cause a potential collision. Following \cite{HLOR11}, we call those cars \emph{helper cars}.

In urban traffic, a helper car for the ego car either has an own reservation on at least one crossing segment of the considered intersection or is approaching it from any direction. The case where a car is driving on a crossing segment is formalised by the \emph{on crossing check}
      \begin{align}\label{formula:oc}
	      oc(c) \; \equiv \; \langle re(c) \wedge cs \rangle \text{.}
      \end{align}
For the second case, we first introduce the abbreviation \emph{one lane} 
    \begin{align*}
	ol\;\equiv\; (true \chop\free\chop true) \vee \exists c: (re(c) \vee cl(c))\text{,}
    \end{align*}
stating, that there is \emph{exactly one lane} occupied with something. While tempting, it is not sufficient to use only $true$ instead of $ol$ because the formula $true$ also holds for zero lanes. If the ego car is approaching an intersection within the distance $d_c$, its crossing controller is supposed to start claiming crossing segments. For an arbitrary other car $C$ approaching the intersection from the opposite side of the intersection, we do not know the braking distance and therefore add the maximum safety envelope $se\;max(C)$ to $d_c$, yielding the distance $d_{c}' = d_c + max\;se(C)$. We formalise that a car approaches an intersection from the opposite side of the intersection within the distance $d_{c}'$ with the \emph{opposing car approaching the crossing check}
\begin{align}\label{formula:ocac}
	ocac(c) \;\equiv\; \left\langle {\begin{array}{c} ol \\ re(\ego) \end{array}} \right\rangle \chop \left\langle {\begin{array}{c} cs \chop \neg \langle cs \rangle \wedge {\free}^{< d_{c}'} \chop re(c) \\ ol \end{array}} \right\rangle \wedge dir(c) \text{.}
\end{align}
The atom $dir(c)$ states whether a car drives in the direction of its lane or not, which the ego car is able to perceive with its sensors. This atom is needed to exclude the special case, that $ocac(c)$ comprises a car spatially driving on the requested lane but driving away from the intersection. A car that is driving away from the intersection is not of interest, as its own braking distance can not stretch to the intersection and as it might leave the view of ego car $E$ soon anyway.

We generally forbid a car entering an intersection while changing lanes as the directed edges in our topology do not allow this (cf. Sect.~\ref{sec:topology}). Therefore, we introduce the \emph{lane change check}
    \begin{align}\label{formula:lc}
	lc(c) \; \equiv \; \left\langle {\begin{array}{c} re(c) \\ re(c) \end{array}} \right\rangle \text{.}
    \end{align}

With formulae \eqref{formula:oc}, \eqref{formula:ocac} and \eqref{formula:lc}, the ego car identifies all described suitable helper cars with the \emph{potential helper check}
\begin{align}\label{formula:potentialhelpercheck}
	ph(c) \; \equiv \; c \neq \ego \wedge (oc(c) \vee ocac(c)) \wedge \neg lc(c) \text{.}
\end{align}

\subsection{Crossing Controller}\label{sec:threecontrollers}
\begin{figure}[htbp]
\centering
	\begin{tikzpicture}[initial text=,->,>=stealth',shorten >=1pt,auto,inner sep=1.5pt,minimum size=20pt,node distance=8 cm, semithick, scale = 1.0, transform shape]
        \tikzstyle{state}=[draw, shape=ellipse]
	\begin{scope}
		\node [state, initial]	(q0) [label=above left:]{{\Blue $q_0:$} Safe};
		\node [state]		(q1) [xshift=+0.0cm, right of=q0, label=above right:]{{\Blue $(q_1,q_2):$} Crossing ahead};
		\node [state]		(q2) [yshift=+5.7cm, below of=q1, label=above right:]{{\Blue $q_3:$} \parbox{3cm}{\centering{Wait for} \\ \centering{communication}}};
		\node [state]		(q3) [xshift=0.0cm, yshift=+5.7cm, below of=q0, label=above right:]{{\Blue $(q_4,q_5):$} On crossing};

	    	\path	(q0) edge [bend left=00]  node[above, xshift=+0.0cm, yshift=-0.55cm] {\parbox{3cm}{\centering{approaching} \\ \centering{crossing}}} (q1)
	    		(q1) edge [bend right=00]  node[left, xshift=-0.35cm, yshift=-0.3] {no helper} (q3)
	    		(q1) edge [bend left=20]  node[right, xshift=-0.1cm, yshift=-0.1cm] {\parbox{2.4cm}{\centering{(at least one)} \\ \centering{helper exists}}} (q2)
	    		(q2) edge [bend left=00]  node[above, xshift=+0.0cm, yshift=-0.15cm] {all \emph{yes}} (q3)
	    		(q2) edge [bend left=20]  node[left, xshift=+0.1cm, yshift=-0.1cm] {\parbox{1.7cm}{\centering{one \emph{no} or} \\ \centering{timeout}}} (q1)
	    		(q3) edge [bend left=00]  node[left, xshift=-0.0cm, yshift=+0.0cm] {finished} (q0);
	\end{scope}
	\end{tikzpicture}
	\caption[]{Overview over crossing controller protocol.}
	\label{fig:ccoverview}
\end{figure}
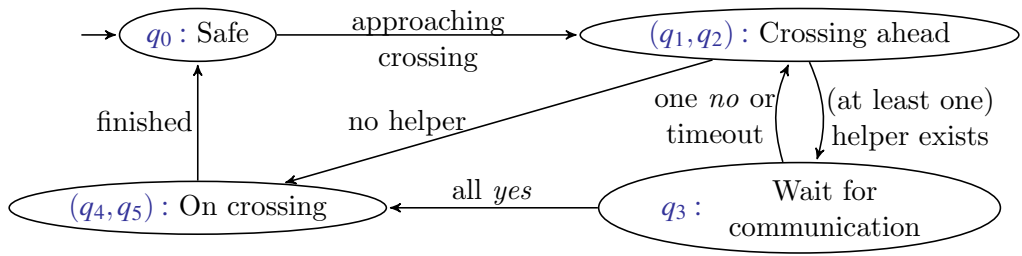

We now construct the \cc{} \crp{} for turning manoeuvres on crossings with imperfect knowledge. The overall goal of the crossing controller is to perform turn manoeuvres at intersections while always maintaining the safety property \eqref{formula:safe}. A coarser version of the detailed \cc{} \crp{} depicted in Fig.~\ref{fig:crossingcontroller} is shown in Fig.~\ref{fig:ccoverview}.

\noindent
\textbf{Overview (cf. Fig.~\ref{fig:ccoverview}).} We assume the initial state of the controller to be Safe, i.e. no collision exists. When a crossing is ahead, the car may enter the intersection by itself, iff no helper exists (e.g. the multi-view is empty except for the ego car). If at least one potential helper exists, the actor needs to communicate with the helpers. If one helper sends a \emph{no}-message or one helper does not answer, the actor withdraws the crossing claim and may try to enter the intersection later again (somewhen the conflicting other car will have left the intersection). Iff all helpers send a \emph{yes}-message, the ego car can safely enter the intersection and finish the crossing manoeuvre.

\noindent
\textbf{Details (cf. Fig.~\ref{fig:crossingcontroller}).} We introduce a \emph{collision check} $col(\ego)$ whose negation $\neg col(\ego)$ holds invariantly in the initial state of our \cc{} and is expressed by the \UMLSL{} formula
    \begin{align}\label{formula:col}
	col (\ego) \; \equiv \; \exists c : c \neq \ego \wedge \langle re(\ego) \wedge re(c) \rangle \text{.}
    \end{align}
The crossing controller only becomes active and leaves its initial safe state, if $E$ approaches an intersection within less than the previously introduced distance $d_c$ with no other car between the actor and the intersection. For this, we formalise the \emph{crossing ahead check}
    \begin{align}\label{formula:ca}
	ca(\ego) \; \equiv \; \langle re(\ego) \chop {\free}^{< d_c} \wedge \neg \langle cs \rangle \chop cs \rangle \text{.}
    \end{align}
The crossing controller \emph{claims} the crossing segments needed for the turn manoeuvre with the controller action \texttt{cc(}$ego$\texttt{)}. Then it checks for a potential collision $pc(c)$ \eqref{formula:pc} with an arbitrary car $c$ and possibly withdraws the crossing claim. Else with the potential helper check $ph(c)$ \eqref{formula:potentialhelpercheck} it evaluates if a helper for the manoeuvre is available, where we observe two possible results:
\begin{compactenum}
    \item No helper car is available or
    \item At least one helper car exists.
\end{compactenum}
In the first case, the controller proceeds without help. If $lc(\ego)$ \eqref{formula:lc} and $pc(c)$ \eqref{formula:pc} do not hold, the actor reserves the claimed crossing segments and starts the crossing manoeuvre. To prevent deadlocks, we set a time bound $t_{o}$ for the time that may pass between claiming and reserving crossing segments. If the actor reserves crossing segments, the on crossing check $oc(\ego)$ holds invariantly. We assume a crossing manoeuvre to take at most $t_{cr}$ time to finish. Once the actor has left the last crossing segment and is driving on a lane, the crossing manoeuvre is finished. The reservation of $E$ is then reduced to the next segment after the intersection in $\pth(E)$.

If helper cars are available, the crossing controller needs to communicate because of the missing information about the braking distances of the helpers. $E$ sends the output message $cross!\langle ego, cs \rangle$, where $cs$ is the set of crossing segments the ego car claims according to $\pth(E)$. If $E$ receives its own car identifier via channel $no$, it immediately withdraws its claim and changes back to $q_1$. While only one \emph{no}-message is sufficient to abort the crossing manoeuvre, it is not enough to receive only one \emph{yes}-message. Therefore, the controller waits $t_w$ time units for the answers of the helpers, where we assume $t_w$ to be a worst case time bound in which all helpers are technically able to answer. For realistic worst case time bounds in real-time broadcast communication, we e.g. refer to the work of Asplund et al.~\cite{An12}.

$E$ collects all identifiers of helpers that answered via channel $yes$ in a set $\mathbb{H}$. After $t_w$ time, it compares $\mathbb{H}$ with the available potential helpers with $ \neg \exists c \in \mathbb{I} \backslash \mathbb{H} : ph(c)$. Then it either reserves the claimed crossing segments, or withdraws the claim, if at least one potential helper did not answer. Once the crossing controller entered state $q_3$ and thus started the communication, it informs the helpers when it either withdraws a claim or successfully finishes the manoeuvre via broadcast channel $finish$. The constructed crossing controller is depicted in Fig.~\ref{fig:crossingcontroller}.

\begin{figure}[htbp]
\centering
	\begin{tikzpicture}[initial text=,->,>=stealth',shorten >=1pt,auto,node distance=4 cm,
                semithick, inner sep=1.0pt,minimum size=20pt, scale = 1.0, transform shape]
        \tikzstyle{state}=[draw, shape=ellipse]
	\begin{scope}
		\node [state, initial]	(q0) [label=above left:]{{\Blue $q_0 :$} $\neg col(\ego)$};
		\node [state]		(q1) [xshift=+0.2cm, right of=q0, label=above right:]{{\Blue $q_1 :$} $ca(\ego)$};
		\node [state]		(q2) [xshift=+2.0cm, right of=q1, label=above right:]{{\Blue $q_2:$} \parbox{1.56cm} {\centering{$ca(\ego)$} \\ \centering{$\wedge\; t \leq t_o$}}};
		\node [state]		(q3) [xshift=-0.0cm, yshift=+0.1cm, below of=q2, label=above right:]{{\Blue$q_3:$} \parbox{2.0cm}{\centering{$ca(\ego)$} \\ \centering{$\wedge \neg \exists c: pc(c)$} \\ \centering{$\wedge\; x \leq t_{w}$}}};
		\node [state]		(q4) [yshift=+0.1cm, below of=q0, label=above right:]{{\Blue $q_4:$} \parbox{1.56cm} {\centering{$x \leq t_{cr}$} \\ \centering{$\wedge\; oc(\ego)$}}};
		\node [state]		(q5) [yshift=-1.7cm, above of=q2, label=above right:]{{\Blue $q_5:$} \parbox{1.56cm} {\centering{$x \leq t_{cr}$} \\ \centering{$\wedge\; oc(\ego)$}}};		

	    	\path	(q0) edge [bend left=00]  node[above, xshift=+0.0cm, yshift=+0.1cm] {$ca(\ego)$} (q1)
	    		(q1) edge [bend left=5]  node[above, yshift=-0.15cm] {\texttt{cc(}$\ego$\texttt{)}$;x:=0$} (q2)
	    		(q2) edge [bend left=5]  node[below, xshift=+0.0cm, yshift=+0.05cm] {$\exists c : pc(c) /$ \texttt{wd cc(}$\ego$\texttt{)}} (q1)
	    		(q2) edge [bend left=00]  node[left, xshift=+0.4cm, yshift=+0.1cm] {\parbox{6cm}{\centering{$\neg \exists c: (pc(c) \vee ph(c)) \wedge \neg lc(\ego)$} \\ \centering{$/$ \texttt{rc(}$\ego$\texttt{)}$; x:=0$}}} (q5)
	    		(q2) edge [bend left=00]  node[right, xshift=-0.0cm, yshift=-0.04cm] {\parbox{2.6cm}{$\exists c: ph(c)$ \\ $\wedge \neg \exists c: pc(c)$ \\ $\wedge \neg lc(\ego)$ \\ $/ cross!\langle ego, cs \rangle;$ \\ $\mathbb{H} := \emptyset ; x:=0$}} (q3)
	    		(q3) edge [loop below, min distance=11mm, looseness=-5]  node[left, xshift=+0.15cm, yshift=+0.4cm] {\parbox{4cm}{\centering{$yes?\langle c, d \rangle : c = ego$} \\ \centering{$/ \mathbb{H} := \mathbb{H} \oplus \{d\}$}}} (q3)
	    		(q3) edge [bend left=20]  node[left, xshift=-0.5cm, yshift=+0.1cm] {\parbox{4.5cm}{\centering{$no?c : c = ego$} \\ \centering{$\vee (x \geq t_w \wedge \exists c \in \mathbb{I} \backslash \mathbb{H} : ph(c))$} \\ \centering{$/$ \texttt{wd cc(}$\ego$\texttt{)}$;$ \\ \centering{$finished!ego$}}}} (q1)
	    		(q3) edge [bend left=00]  node[below, xshift=-0.5cm, yshift=+0.0cm] {\parbox{4.5cm}{\centering{$x\geq t_w \wedge \neg \exists c \in \mathbb{I} \backslash \mathbb{H} : ph(c)$} \\ \centering{$\wedge \neg \exists c: pc(c) \wedge \neg lc(\ego)$} \\ \centering{$/$ \texttt{rc(}$\ego$\texttt{)}$; x:=0$}}} (q4)
	    		(q4) edge [bend left=00]  node[left, xshift=+0.15cm, yshift=+0.0cm] {\parbox{2.2cm}{$x \geq t_{cr}/$ \\ \texttt{wd rc(}$\ego$\texttt{)}$;$ \\ $finished!ego$}} (q0)
	    		(q5) edge [bend right=20]  node[left, xshift=+0.05cm, yshift=+0.0cm] {\parbox{3cm}{$x \geq t_{cr}/$ \\ \texttt{wd rc(}$\ego$\texttt{)}}} (q0);
	\end{scope}
	\end{tikzpicture}
	\caption[Crossing controller \crp]{Crossing controller \crp}
	\label{fig:crossingcontroller}
\end{figure}

\subsection{Helper Cars and Helper Controller}\label{sec:helpercontroller}
\begin{figure}[htbp]
\centering
	\begin{tikzpicture}[initial text=,->,>=stealth',shorten >=1pt,auto,inner sep=1.3pt,minimum size=20pt,node distance=8 cm, semithick, scale = 1.0, transform shape]
        \tikzstyle{state}=[draw, shape=ellipse]
	\begin{scope}
		\node [state, initial]	(q0) [label=above left:]{{\Orange $q_0 :$} idle};
		\node [state]		(q1) [xshift=+0.0cm, right of=q0, label=above right:]{{\Orange $(q_2,q_4):$} Helping};
		\node [state]		(q2) [xshift=-4.1
		cm, yshift=+5.5cm, below of=q1, label=above right:]{{\Orange $(q_1, q_3, q_5):$} Decline requests};

	    	\path	(q0) edge [bend left=5]  node[above, xshift=-0.0cm, yshift=-0.15cm] {\parbox{5.5cm}{\centering{initial request $\&$ no conflict}}} (q1)
	    		(q1) edge [bend left=5]  node[below, xshift=+0.2cm, yshift=+0.1cm] {\parbox{2.7cm}{\centering{helping finished}}} (q0)
			
			(q1) edge [bend left=15]  node[right, xshift=+0.4cm, yshift=-0.0] {\parbox{3.1cm}{\centering{additional request} \\ \centering{$\&$ conflict with}\\\centering{initial enquirer}}} (q2)
	    		(q2) edge [bend left=10]  node[left, xshift=-0.9cm, yshift=-0.4cm] {\parbox{1.5cm}{\centering{request} \\ \centering{declined}}} (q1)

	    		(q0) edge [bend right=15]  node[left, xshift=-0.1cm, yshift=-0.15cm] {\parbox{2.5cm}{\centering{initial request} \\ \centering{but conflict}}} (q2)
	    		(q2) edge [bend right=10]  node[right, xshift=+0.45cm, yshift=-0.4cm] {} (q0);
	\end{scope}
	\end{tikzpicture}
	\caption[]{Overview over helper controller protocol.}
	\label{fig:hcoverview}
\end{figure}
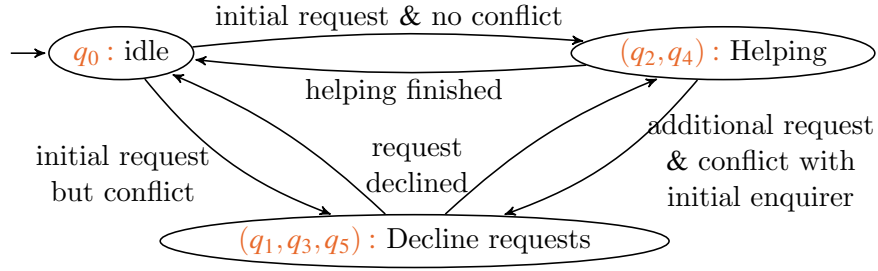

As introduced in Sect.~\ref{sec:imperfectknowledge}, a helper car is either driving on the crossing or approaching it from a different direction than the ego car. An arbitrary car is allowed to be helper for more than one requesting car, e.g. needed if four cars turn simultaneously right at an intersection. We therefore assume that every car owns several clones of the helper controller, but only one of the helper controllers assist one specific car at once. A coarser version of the detailed \hcp{} \hc{} depicted in Fig.~\ref{fig:helpercontroller} is shown in Fig.~\ref{fig:hcoverview}.

\noindent
\textbf{Overview (cf. Fig.~\ref{fig:hcoverview}).} Whenever an idle helper controller receives a crossing request it checks if it meets the helper requirements (on crossing or approaching crossing) and if there exist no potential collision of the request with its own crossing claim or reservation. Then it either declines the request or starts to help the enquirer. If the helper controller receives a conflicting request from another car during the helping process, it declines this request immediately.

\noindent
\textbf{Details (cf. Fig.~\ref{fig:helpercontroller}).} In the helper controller we use the unique variable $a$ to identify the helper controller and we call a car searching for a helper \emph{enquirer} or \emph{enquiring car}. The set $cs_a := cclm(a) \cup cres(a)$ denotes the claimed and reserved crossing segments of the helper car. If a car receives a broadcast request $cross![c,cs]$, its \hcp{} first checks if it is a potential helper for $c$ with the \emph{inverse potential helper check}
\begin{align}\label{formula:invph}
	ph^{-1}(c,cs) \; \equiv \; a\neq c \wedge (oc(a) \vee ca(a)) \wedge \neg lc(a) \wedge (cs_a \cap cs = \emptyset) \text{.}
\end{align}
With the first part of the formula, the potential helper checks if its position is suitable and whether it is currently changing lanes. With the latter part of the formula the potential helper checks for disjointedness of its own segments $cs_a$ and the received crossing segments $cs$. Note that this check resembles the potential collision check for lanes. If the controller detects a potential collision, it immediately sends a \emph{no}-message to the enquiring car.

If it is a potential helper, the controller sends a \emph{yes}-message to the enquiring car in less than $t$ time units, where $t < t_w$ and $t_w$ is the time bound the crossing controller waits for the answers of the helpers. While helping, it additionally declines crossing requests from a third car whose request overlaps with the crossing segments of the car the helper already assists. If the helper left the intersection or if the crossing manoeuvre of the enquirer is finished, the helping process is finished. The resulting helper controller is depicted in Fig.~\ref{fig:helpercontroller}.

\begin{figure}[htbp]
\centering
	\begin{tikzpicture}[initial text=,->,>=stealth',shorten >=1pt,auto,node distance=5.0cm, semithick, inner sep=1.0pt,minimum size=20pt, scale = 1.0, transform shape]
        \tikzstyle{state}=[draw, shape=ellipse]
	\begin{scope}
		\node [state, initial]	(q0) [label=above left:]{\Orange$q_0$};
		\node [state]		(q1) [yshift=+1.8cm, below of=q0, label=above right:]{{\Orange$q_1 :$}$\mathbf{U}$};
		\node [state]		(q2) [xshift=+1.1cm, right of=q0, label=above right:]{{\Orange$q_2 :$}{\parbox{1.9cm}{\centering{$ph^{-1}(h, {cs}_h)$}\\ \centering{$\wedge x < t$}}}};
		\node [state]		(q3) [yshift=-2.5cm, yshift=-0.0cm, above of=q2, label=above right:]{{\Orange$q_3:$}$\mathbf{U}$};
		\node [state]		(q4) [xshift=+1.3cm, right of=q2, label=above right:]{{\Orange$q_4:$}{\parbox{1.9cm}{\centering{$ph^{-1}(h, {cs}_h)$}\\\centering{$\wedge x \leq t_w + t_{cr}$}}}};
		\node [state]		(q5) [yshift=-2.5cm, above of=q4, label=above right:]{{\Orange$q_5:$}$\mathbf{U}$};
		
	    	\path	(q0) edge [bend left=00] node[right, xshift=-0.45cm, yshift=-1.1cm] {\parbox{4.0cm}{\centering{$cross?\langle c, cs \rangle:c \neq a$} \\\centering{$\wedge cs \cap cs_{a} \neq \emptyset$}\\\centering{$/ d:=c$}}} (q1)
			(q1) edge [bend left=20] node[left, xshift=+0.0cm, yshift=+0.0cm] {$/no!d$} (q0)
			
			(q0) edge [bend left=00] node[above, xshift=+0.0cm, yshift=+0.0cm] {\parbox{5cm}{\centering{$cross?\langle c, cs \rangle: ph^{-1}(c, cs)$} \\ \centering{$/ h:=c; {cs}_h:=cs; x:=0$}}} (q2)
			(q2) edge [bend left=15] node[below, xshift=+1.7cm, yshift=-0.05cm] {\parbox{5cm}{\centering{$finished?c: c=h \vee x \geq t$} \\ \centering{$\vee \neg ph^{-1}(h, {cs}_h) /no!h$}}} (q0)
			
			(q2) edge [bend left=00] node[left, xshift=+0.1cm, yshift=+0.2cm] {\parbox{3.5cm}{\centering{$cross?\langle c, cs \rangle: c \neq h$} \\ \centering{$\wedge cs_h \cap cs \neq \emptyset / d:=c$}}} (q3)
			(q3) edge [bend left=20] node[right, xshift=+0.0cm, yshift=+0.0cm] {$no!d$} (q2)
			
			(q2) edge [bend left=00] node[above, xshift=+0.0cm, yshift=+0.0cm] {\parbox{5cm}{\centering{$ph^{-1}(h, {cs}_h) \wedge x < t$} \\ \centering{$/yes!\langle h , a \rangle; x:=0$}}} (q4)
			
			(q4) edge [bend left=00] node[left, xshift=+0.1cm, yshift=+0.2cm] {\parbox{3.5cm}{\centering{$cross?\langle c, cs \rangle: c \neq h$} \\ \centering{$\wedge {cs}_h \cap cs \neq \emptyset / d:=c$}}} (q5)
			(q5) edge [bend left=20] node[right, xshift=+0.0cm, yshift=+0.0cm] {$no!d$} (q4)
			
			(q4) edge [bend left=35] node[below, xshift=+0.5cm, yshift=-0.05cm] {\parbox{5cm}{\centering{$finished?c: c=h$} \\\centering{$\vee \neg ph^{-1}(h, {cs}_h) \vee x \geq t_{cr} + t_w$}}} (q0);
	\end{scope}
	\end{tikzpicture}
	\caption[Helper controller \hc]{Helper controller \hc}
	\label{fig:helpercontroller}
\end{figure}

\section{Conclusion}\label{sec:conclusion}
We extend our approach for urban traffic manoeuvres with perfect knowledge from \cite{HS16} by a more realistic concept of imperfect knowledge, where autonomous cars have no information about speed and braking distances of other cars. To this end, we introduce a multi-view semantics for \UMLSL{} formulae. We propose broadcast communication with data constraints to specify our communicating crossing controllers, which can autonomously perform turn manoeuvres at intersections with the help of controllers in other cars at the intersection.

\noindent
\textit{More on related work.} Linker \cite{L15} and Ody \cite{O15} present undecidability results of the spatial part of \MLSL, which unfortunately apply for our extension \UMLSL, too. However, Fränzle et al. \cite{FHO15} prove that \MLSL{} is decidable, when considering only a bounded \emph{scope} around the cars. This is a constraint motivated by reality because actual autonomous cars can only process state information of finitely many environmental cars in real-time.

\noindent
\textit{Future work.} The purely formal specification of our controllers, detached from the car dynamics, allows for formal verification of the safety condition \eqref{formula:safe} from p.~\pageref{formula:safe} as future work. The proof idea is as follows: we show safety \eqref{formula:safe} from the viewpoint of an arbitrary actor $E$ with that approaches an intersection and thus generates a multi-view $V_m (E)$ (cf. Sect.~\ref{sec:view}). We assume an initial safe \traffics{} ${\Road}_0$ and inductively show for every \traffics{} ${\Road}_{k}$, reachable from ${\Road}_0$ by $k$ evolution transitions, that it is also safe. For this purpose, we propose to separate the proof of spatial properties in \UMLSL{} guards and invariants in the controllers from the proof for their timing and communication behaviour. The spatial part can be shown either by exploiting directly the semantics of guards and invariants in the controllers or by using an adaptation of the proof system introduced for standard \MLSL{} in \cite{L15}. For the time and communication part of the extended timed automata controllers, we aim for a proof with assistance of UPPAAL \cite{BDL04}.

The here proposed crossing controller is safe, but not deadlock free, wherefore it is interesting to examine a \emph{(timed) liveness} property. By extending \UMLSL{} with operators from Koymans \emph{metric temporal logic} \cite{K90}, we could express, that a car approaching an intersection ($ca(c)$) and that desires to cross it (${pth(c)}_{next(c)} \in \CS$), \emph{finally} ($\mathbf{F}$) passes it in less than $t$ time units ($<t$):
\begin{align*}
  \mathit{Life} \; \equiv \; \forall c: (ca(c) \wedge {pth(c)}_{next(c)} \in \CS \rightarrow {\mathbf{F}}_{<t} oc (c))\text{.}
\end{align*}
The relation of our work to game theoretical approaches is interesting. We could e.g. use UPPAAL TiGa \cite{C05} for our purposes, where an extended timed automaton represents two players: the system itself and the environment. As environmental part, we could model the time out transitions of our controllers. The systems' goal is to reach a specific state (e.g. a state where \emph{on crossing} ($oc(c)$) holds invariantly) or avoid a specific state (e.g. a \emph{bad state} with a time out).

For now, we conveniently assumed broadcast communication as we already used it in previous approaches. For future work it is interesting to link our communication requirements more detailed to communication standards from Car2Car Communication (cf. Kenney \cite{K11}).

\bibliographystyle{eptcs}
\bibliography{bib}
\end{document}